\documentclass[a4paper]{jpconf}

\usepackage{amsmath}
\usepackage{amssymb}
\usepackage{epsfig}
\usepackage{graphicx}
\usepackage{multirow}
\usepackage{color}
\usepackage[normal]{subfigure}
\usepackage{rotating}

\newcommand{\capdef}{}
\newcommand{\mycaption}[2][\capdef]{\renewcommand{\capdef}{#2}
\caption[#1]{{\footnotesize #2}}}

\newcommand{\bwt}{\begin{widetext}}
\newcommand{\ewt}{\end{widetext}}
\newcommand{\be}{\begin{equation}}
\newcommand{\ee}{\end{equation}}
\newcommand{\bdm}{\begin{displaymath}}
\newcommand{\edm}{\end{displaymath}}
\newcommand{\bea}{\begin{eqnarray}}
\newcommand{\eea}{\end{eqnarray}}
\newcommand{\nn}{\nonumber}

\def\eq#1{{Eq.~(\ref{#1})}}
\def\eqs#1#2{{Eqs.~(\ref{#1})--(\ref{#2})}}
\def\fig#1{{Fig.~\ref{#1}}}

\def\vev#1{\left\langle #1 \right\rangle}

\def\Tr{\mbox{Tr}\,}

\begin{document}
\title{Aspects of symmetry breaking in $SO(10)$ GUTs}

\author{Luca Di Luzio\footnote{In collaboration with Stefano Bertolini (INFN and SISSA, Trieste) and Michal Malinsk\'y (IFIC, Valencia).
Talk given at DISCRETE'10 - \emph{Symposium on Prospects in the Physics of Discrete Symmetries}, 6-11 Dec 2010, Rome.
}}

\address{SISSA and INFN, Sezione di Trieste,
Via Bonomea 265, 34136 Trieste, Italy}

\ead{diluzio@sissa.it}

\begin{abstract}
I review some recent results on the Higgs sector of minimal $SO(10)$ grand unified theories both with and without supersymmetry. It is shown that nonsupersymmetric $SO(10)$ with just one adjoint triggering the first stage of the symmetry breaking does provide a successful gauge unification when radiative corrections are taken into account in the scalar potential, while in the supersymmetric case it is argued that the troubles in achieving a phenomenologically viable breaking with representations up to the adjoint are overcome by considering the flipped $SO(10)$ embedding of the hypercharge.
\end{abstract}

%%%%%%%%%%%%%%%%%%%%%%%%%%%%%%%%%%%%%%%%%%%
\section{Introduction}
%%%%%%%%%%%%%%%%%%%%%%%%%%%%%%%%%%%%%%%%%%%

While the standard model (SM) matter quantum numbers nicely fit in a few lowest-dimensional representations of the unified groups such as $SU(5)$ or $SO(10)$, 
this synthetic process has no counterpart in the Higgs sector where a larger set 
of higher-dimensional representations 
is usually needed in order to spontaneously break the enhanced gauge symmetry down to the SM gauge group. 
In this respect, establishing the minimal Higgs content needed for the grand unified theory (GUT) breaking is a basic question 
which has been already addressed in the early days of the GUT program.
Let us remark that the quest for the simplest Higgs sector is driven not only by aesthetic criteria but it is also a phenomenologically relevant issue related to tractability and predictivity of the models.  
Indeed, the details of the symmetry breaking pattern, sometimes overlooked in the phenomenological analysis, give further constraints on the low-energy observables such as the proton decay and the effective SM flavor structure.

Here we will focus mainly on $SO(10)$ GUTs, both with and without supersymmetry (SUSY). 
Let us recall that before considering any symmetry breaking dynamics the following representations are required by the group 
theory in order to achieve a full breaking of $SO(10)$ to the SM:
\begin{enumerate}
\item $16_H$ or $126_H$ for rank reduction. 
Their SM-singlet vacuum expectation value (VEV) communicates the $B-L$ breaking to neutrino masses, but preserves an $SU(5)$ little group. 
\item $45_H$ or $54_H$ or $210_H$ for the further $SU(5)$ breaking. 
They admit for little groups different from $SU(5) \otimes U(1)$, yielding the SM when intersected with the $SU(5)$ remnant of (i).
\end{enumerate}
It should be also mentioned that a one-step $SO(10) \rightarrow \text{SM}$ breaking can be achieved via only one $144_H$ irreducible Higgs representation \cite{Babu:2005gx}. 
However, such a setting requires an extended matter sector, including $45_F$ and $120_F$ multiplets, in order to accommodate realistic fermion masses \cite{Nath:2009nf}.
While the choice between $16_H$ or $126_H$ is a model dependent issue related to the details of the Yukawa sector, 
the simplest option in the list (ii) is certainly given by the adjoint $45_H$. However, since the early 1980's, it has been observed that the vacuum dynamics aligns the adjoint along 
an $SU(5) \otimes U(1)$ direction, making the choice of the $45_H$ alone not phenomenologically viable. 
In the nonsupersymmetric case the alignment is only approximate \cite{Yasue:1980fy,Anastaze:1983zk,Babu:1984mz},
but it is such to clash with unification constraints 
which do not allow for any $SU(5)$-like intermediate stage, while in the supersymmetric limit 
the alignment is exact due to F-flatness \cite{Buccella:1981ib,Babu:1994dc,Aulakh:2000sn}, thus never landing to a supersymmetric SM vacuum. 
In the next sections we will review all these issues in more detail and provide a way out 
to the aforementioned problem of the vacuum alignment.

%%%%%%%%%%%%%%%%%%%%%%%%%%%%%%%%%%%%%%%%%%%
\section{Adjoint breaking in nonsupersymmetric $SO(10)$}
%%%%%%%%%%%%%%%%%%%%%%%%%%%%%%%%%%%%%%%%%%%
Let us consider the most general renormalizable tree level scalar potential which
can be constructed out of $45_H$ and $16_H$ in nonsupersymmetric $SO(10)$ 
\be
V_0=V_{45_H}+V_{16_H}+V_{45_H 16_H} \, , 
\label{potential4516}
\ee
where, according to the notation in Ref.\! \cite{Bertolini:2009es}
\begin{align}
& V_{45_H}=
-\frac{\mu^2}{2}\, \Tr 45_H^2 + \frac{a_1}{4}\, (\Tr 45_H^2)^2 + \frac{a_2}{4}\, \Tr 45_H^4 \, , \\
& V_{16_H}=
-\frac{\nu^2}{2}\, 16_H^\dag 16_H
+\frac{\lambda_1}{4}\, (16_H^\dag 16_H)^2
 +\frac{\lambda_2}{4}\, (16_H^T \, \Gamma \, 16_H)(16_H^\dag \, \Gamma \, 16_H^\ast) \, , \\
& V_{45_H 16_H}=
\alpha\, (16_H^\dag 16_H)\Tr 45_H^2+\beta\, 16_H^\dag 45_H^2 16_H
+\tau\, 16_H^\dag 45_H 16_H \, .
\end{align}
There are in general three SM-singlets in the $45_H\oplus16_H$ reducible representation of $SO(10)$.
Their VEVs are defined in the following way 
\be
\omega_{B-L} \subset \vev{(15,1,1)} \subset \vev{45_H} \, , \quad
\omega_{R} \subset \vev{(1,1,3)} \subset \vev{45_H} \, , \quad
\chi_{R} \subset \vev{(\overline{4},1,2)} \subset \vev{16_H} \, ,
\ee
where the three submultiplets above  
are labeled according to the $SO(10)$ subalgebra
$4_{C}\, 2_{L}\, 2_{R}$ (shorthand notation for $SU(4)_C \otimes SU(2)_L \otimes SU(2)_R$).
Different VEV configurations trigger the spontaneous breakdown of the $SO(10)$ symmetry into the following subgroups. 
Using a self-explanatory notation, for $\chi_{R}= 0$ one finds
\begin{align}
\label{vacua}
&\omega_{R}= 0,\, \omega_{B-L}\neq 0\; : & 3_c\, 2_L\, 2_R\, 1_{B-L} \, , \nn \\
&\omega_{R}\neq 0,\, \omega_{B-L}= 0\; : & 4_{C} 2_L 1_R \, , \nn \\
&\omega_{R}\neq 0,\, \omega_{B-L}\neq 0\; : & 3_c\, 2_L\, 1_R\, 1_{B-L} \, ,   \\
&\omega_{R}=-\omega_{B-L}\neq 0\; : & \mbox{flipped}\, SU(5) \otimes U(1)_Z \, , \nn \\
&\omega_{R}=\omega_{B-L}\neq 0\; :  & \mbox{standard}\, SU(5) \otimes U(1)_Z \, . \nn
\end{align}
When $\chi_{R}\neq 0$ all intermediate symmetries are spontaneously broken down to the 
$3_{c}\,2_{L}\,1_{Y}$ of the SM with the exception of the last case which leaves the standard $SU(5)$ unbroken. 
In this language, the potentially viable breaking chains fulfilling the basic gauge unification constraints (labeled as VIII and XII in Ref.\! \cite{Bertolini:2009qj}) correspond to the settings with:
\begin{eqnarray}
&& \omega_{B-L}\gg \omega_{R} > \chi_{R}\;:\;  SO(10)\to 3_{c}2_{L}2_{R}1_{B-L}\to 3_{c}2_{L}1_{R}1_{B-L}\to 3_{c}2_{L}1_{Y} \qquad\; \text{Chain VIII} \, , \nn\\
&& \omega_{R}\gg \omega_{B-L} > \chi_{R}\;:\;  SO(10)\to 4_{C}2_{L}1_{R}\to 3_{c}2_{L}1_{R}1_{B-L}\to 3_{c}2_{L}1_{Y} \qquad\qquad \text{Chain XII} \, . \nn
\end{eqnarray}
The parameters (couplings and VEVs) of the scalar potential are constrained by the requirements
of boundedness and the absence of tachyonic states, ensuring that
the vacuum is stable and the stationary points correspond to physical minima.
In particular, from the shape of the tree level masses of the $(8,1,0)$ and $(1,3,0)$ SM sub-multiplets of $45_{H}$:  
\bea
\label{310PGBmass}
M^2(1,3,0) & = & 
2 a_2 (\omega _{B-L} - \omega _R) (\omega _{B-L} + 2 \omega _R) \, , \\
\label{810PGBmass}
M^2(8,1,0) & = & 
2 a_2 (\omega _R - \omega _{B-L}) (\omega _R + 2 \omega _{B-L}) \,,
\eea
which can not be simultaneously positive unless
\be
\label{bound130810}
a_2 < 0 \, , \qquad -2<\omega_{B-L}/\omega_R<-\tfrac{1}{2} \,,
\ee
one concludes that the only vacuum configurations allowed are those in the close vicinity of 
the flipped $SU(5) \otimes U(1)_Z$ setting.
Hence, the large
hierarchy
(of about four orders of magnitude)
between $\omega_R$ and $\omega_{B-L}$, required 
by gauge coupling unification (cf.\! chains VIII and XII in Ref.\! \cite{Bertolini:2009qj}) cannot be achieved.
This is the key point of the classical argument that the nonsupersymmetric $SO(10)$ GUTs with only one adjoint responsible for the first stage of the $SO(10)$ breakdown can not support the phenomenologically favoured symmetry breaking chains. 

%=====================================================================
\subsection{A tree level accident}
%=====================================================================

The rationale for understanding the strong correlation among the masses of the states $(1,3,0)$ 
and $(8,1,0)$ can be obtained by looking at the enhancement of the global symmetry in a trivial limit 
of the scalar potential. 
When only trivial invariants of both $45_H$ and $16_H$ are considered 
($a_2 = \lambda_2 = \beta = \tau = 0$) 
the global symmetry of $V_{0}$ is $O(45)\otimes O(32)$.
This is then broken spontaneously into $O(44)\otimes O(31)$ by the $45_H$ and $16_H$ VEVs yielding 44+31=75 Goldstone bosons (GB) in the scalar spectrum.
The gauge $SO(10)$ symmetry is at the same time broken down to the SM gauge group.
Therefore 75-33=42 pseudo-Goldstone bosons (PGB) are left in the spectrum and
their masses should generally receive contributions
from all the explicitly breaking terms $a_2$, $\lambda_2$, $\beta$ and $\tau$.
Since the states $(1,3,0)$ and $(8,1,0)$ belong to this set of PGB, 
generally one would expect their mass to depend on all of $a_2$, $\lambda_2$, $\beta$, $\tau$ parameters.   
The absence of mass contributions proportional to $\lambda_2$, $\beta$, $\tau$ is just an easily understood 
accident of the tree level potential \cite{Bertolini:2009es}, but nothing prevents those couplings from contributing 
to the PGB masses at the quantum level.

%=====================================================================
\subsection{The quantum vacuum}
%=====================================================================

The relevant one-loop correction to the $(1,3,0)$ and $(8,1,0)$ PGB masses can be conveniently 
computed by means of the one-loop effective potential (EP).
The one-loop EP can be formally written as
\begin{equation}
V_{\rm eff}=V_{0}+\Delta V_s+\Delta V_f+\Delta V_g \, ,
\label{Veff}
\end{equation}
where $\Delta V_{s,f,g}$ denote the contributions from scalars, fermions and gauge bosons respectively.
In dimensional regularization with modified minimal subtraction
($\overline{\text{MS}}$) and in the Landau gauge, they are given by
\be
\Delta V_i (45_H ,16_H ,\mu)= 
\frac{\alpha_i }{64\pi^2}
\Tr\left[\mathcal{M}_i^4(45_H ,16_H)\left(\log\frac{\mathcal{M}_i^2(45_H ,16_H)}{\mu^2}
-\beta_i \right)\right] \, ,
\ee
where $(\alpha_s, \alpha_f, \alpha_g) = (1, -2, 3)$, 
$(\beta_s, \beta_f, \beta_g) = (\tfrac{3}{2}, \tfrac{3}{2}, \tfrac{5}{6})$ 
and $\mathcal{M}_s$, $\mathcal{M}_f$ and $\mathcal{M}_g$ are the 
functional scalar, fermion and gauge boson mass matrices respectively, as obtained from the tree level potential. 
In the following we will neglect the fermionic component of the EP since there are no fermionic states at the unification scale $M_G$.
The running masses $\overline{m}_{ab}^2$ are defined by
\be
\overline{m}_{ab}^2 \equiv
\frac{\partial^2 V_{\rm eff}(45_H ,16_H ,\mu)}{\partial\psi_a\partial\psi_b}\Big |_{\vev{\psi}} \, ,
\label{runningmass}
\ee
where $\psi_{a}$ and $\psi_{b}$ are generic scalar field components 
and the VEVs (denoted collectively $\vev{\psi}$) obey the one-loop stationary equations. 
For a given eigenvalue the physical (pole) masses $M_a^2$ are then obtained by 
\be
M_a^2 = \overline{m}_{a}^2
+ \Delta\Sigma_a (M_a^2) \, ,
\label{physmasses}
\ee
where
$\Delta\Sigma_{a}(p^2) = \Sigma_{a}(p^2) - \Sigma_{a}(0)$
and $\Sigma_{a}$ are the $\overline{\mbox{MS}}$ renormalized self-energies.
Of particular interest is the case when $M_a$ is substantially smaller than
the mass ($M_G$) of the particles that contribute to $\Sigma_{a}$.
At $\mu=M_G$ in the limit $M_{a}^2 \ll M_G^2$ one has
\be
\Delta\Sigma_{a}(M_a^2) = O(M_a^4/M_G^2) \, .
\label{DeltaS}
\ee
In this case the running mass computed from \eq{runningmass} contains the leading gauge independent corrections.

The stringent tree level
constraint on the ratio $\omega_{B-L}/\omega_R$, coming from the positivity
of the masses of the states $(1,3,0)$ and $(8,1,0)$, that forbids
non-$SU(5)$ vacua, follows from the fact that the masses
depend only on the parameter $a_2$.
On the other hand, from the discussion of the would-be global symmetries of the scalar potential we should in general expect their masses to depend on other terms in the scalar potential 
(in particular $\beta$, $\tau$ and gauge interactions at the one-loop level).
The calculation of the EP running mass
from \eq{runningmass} leads for the states $(1,3,0)$
and $(8,1,0)$ at $\mu=M_G$ to the mass shifts
\bea
\label{310onthevac}
&& \!\!\!\!\!\!\!\!\!\!\!\!\!\!\!\! \Delta M^2(1,3,0) = \frac{1}{4\pi^2} \left[ \tau^2
+\beta^2(2\omega_R^2-\omega_R\omega_{B-L}+2\omega_{B-L}^2)
+g^4 \left(16 \omega _R^2+ \omega _R \omega _{B-L} +19 \omega _{B-L}^2\right)\right] 
\quad \\
\label{810onthevac}
&& \!\!\!\!\!\!\!\!\!\!\!\!\!\!\!\! \Delta M^2(8,1,0) = \frac{1}{4\pi^2} \left[ \tau^2
+\beta^2(\omega_R^2-\omega_R\omega_{B-L}+3\omega_{B-L}^2)
+g^4 \left(13 \omega _R^2+ \omega _R \omega _{B-L} +22 \omega _{B-L}^2\right)\right] 
\quad
\eea
where the subleading (and gauge dependent) logarithmic terms
are neglected and we have taken for simplicity $\chi_R = 0$,  
given that $\chi_R \ll \omega_{R,B-L}$ by unification constraints. 
For more details we refer the reader to Refs.\! \cite{Bertolini:2009es,Bertolini:2010ng}.
By comparing \eqs{310onthevac}{810onthevac} with \eqs{310PGBmass}{810PGBmass} it is clear that a consistent scalar mass spectrum can be obtained
for the non-$SU(5)$ vacua, at variance with the tree level result. 
In particular, a hierarchy between $\omega_{B-L}$ and $\omega_R$ (as required by unification), 
while keeping the scalar states positive (minimum condition), is now possible just by taking $|a_2| \lesssim 10^{-2}$. This corresponds to keeping the masses of the PGB $(1,3,0)$ and $(8,1,0)$ one order of magnitude below $M_G$, which makes the EP potential computation self-consistent (since the self-energies in \eq{physmasses} can be neglected) and has the welcome effect of raising a little $M_G$.
Let us also stress that this result is inherent to all the nonsupersymmetric $SO(10)$ models with one adjoint $45_H$ triggering the first stage of the GUT breaking: just one additional GUT representation interacting with the adjoint is needed in order to open up the non-$SU(5)$ vacua at the quantum level.

Nevertheless, the simplest scenario featuring the Higgs scalars in $10_H\oplus 16_H\oplus 45_H$ of $SO(10)$ (where the $10_H$ is needed for the electroweak symmetry breaking) fails when addressing the neutrino spectrum: in nonsupersymmetric models the $B-L$ breaking scale, $M_{B-L}$, turns out to be generally a few orders of magnitude below $M_{G}$ \cite{Bertolini:2009qj}. Thus, the scale of the right-handed (RH) neutrino masses $M_R \sim M_{B-L}^{2}/M_P$ emerging first at the $d=5$ level from an operator of the form $16_{F}^{2}(16^\ast_{H})^{2}/M_P$  (with $M_P$ typically identified with the Planck scale) undershoots by many orders of magnitude the range of about $10^{13 \div 14}$ GeV naturally suggested by the seesaw mechanism.
This issue can be alleviated by considering $126_H$ in place of $16_H$ in the Higgs sector, since in such a case the neutrino masses can be generated at the renormalizable level by the term $16_{F}^{2} 126^\ast_H$. This lifts the problematic $M_{B-L}/M_P$ suppression factor inherent to the $d=5$ effective mass and yields $M_{R}\sim M_{B-L}$, that might be, at least in principle, acceptable \cite{WIP}. 
Fermion masses and mixings in nonsupersymmetric $SO(10)$ with $10_H \oplus 126_H$ in the Yukawa sector have been recently reconsidered by the authors of Ref.\! \cite{Bajc:2005zf}, which show the 
potential predictivity of such a class of models.

%%%%%%%%%%%%%%%%%%%%%%%%%%%%%%%%%%%%%%%%%%%
\section{Adjoint breaking in supersymmetric $SO(10)$}
%%%%%%%%%%%%%%%%%%%%%%%%%%%%%%%%%%%%%%%%%%%

By invoking TeV-scale SUSY, the qualitative picture changes a lot for neutrinos. Indeed, the gauge running within the minimal supersymmetric SM (MSSM) prefers $M_{B-L}$ in the proximity of $M_{G}$ and, hence, the Planck-suppressed $d=5$  RH neutrino mass operator {$16_{F}^{2}\overline{16}_{H}^{2}/M_{P}$, available whenever $16_H\oplus \overline{16}_H$ is present in the} Higgs sector, can naturally reproduce the desired scale for $M_{R}$. Let us recall that both $16_H$ as well as  $\overline{16}_H$ are required in order to retain SUSY below the GUT scale. 
It is therefore very interesting to consider the minimal Higgs setting based on 
the lowest-dimensional representations (up to the adjoint). On the other hand, it is well known \cite{Buccella:1981ib,Babu:1994dc,Aulakh:2000sn}
that the relevant superpotential does not support,
at the renormalizable level, a supersymmetric breaking of the $SO(10)$
gauge group to the SM. This is due to the constraints on the vacuum manifold
imposed by the $F$- and $D$-flatness conditions which, apart from linking
the magnitudes of the $SU(5)$-singlet ${16}_H$ and ${\overline{16}_H}$
VEVs, make the the adjoint {VEV $\vev{45_{H}}$}
aligned to {$\vev{16_H\overline{16}_{H}}$}.
As a consequence, an $SU(5)$ subgroup of the initial $SO(10)$ gauge symmetry remains unbroken. 

%=====================================================================
\subsection{Renormalizable vs nonrenormalizable scenarios}
%=====================================================================

The alignment of the adjoint with the spinors can be broken by giving up renormalizability and allowing for effective $M_P$-suppressed operators in the superpotential \cite{Babu:1994dc}. However, this option may be rather problematic since it introduces a delicate interplay between physics at two different scales, $M_{G}\ll M_{P}$, with the consequence of spreading the GUT-scale thresholds over several orders of magnitude below $M_{G}$. In turn this may affect $d=5$ proton decay as well as the MSSM gauge unification and it may also jeopardize the neutrino mass generation in the seesaw scheme (cf.\! Ref.\! \cite{Bertolini:2010yz} for a more detailed account of these effects). Thus, although the Planck-induced operators can provide a key to overcome the {$SU(5)$ lock} of the minimal SUSY $SO(10)$ Higgs model with $16_{H}\oplus \overline{16}_{H}\oplus 45_{H}$, such an
effective scenario is prone to {failure} when addressing the measured proton stability and light neutrino phenomenology.

On the other hand, in the standard $SO(10)$ framework with a Higgs sector
built off {the} lowest-dimensional representations (up to the adjoint), it is not possible to achieve a renormalizable breaking even admitting multiple copies of each type of multiplet. Firstly, with a single $45_{H}$ at play, the little group of the adjoint is $SU(5) \otimes U(1)$ regardless of the number of $16_{H}\oplus \overline{16}_{H}$ pairs.
The same is true with a second $45_{H}$ added into the Higgs sector because there is no renormalizable mixing among the two $45_{H}$'s apart from the mass term that, without loss of generality, can be taken diagonal.
With a third adjoint Higgs representation at play a cubic $45_145_2 45_3$ interaction is allowed.
However, due to the total antisymmetry of the invariant and the fact that the adjoints commute on the SM vacuum,
the cubic term does not contribute to the F-term equations.
From this brief excursus one concludes that 
the $SU(5)$ lock cannot be broken
at the renormalizable level by means of representations up to the adjoint.

Remarkably, all these issues are alleviated if one considers a flipped variant of the supersymmetric $SO(10)$ unification.
What we have shown in Ref.\! \cite{Bertolini:2010yz} is that the flipped $SO(10)\otimes U(1)_X$ scenario \cite{Kephart:1984xj,Rizos:1988jn,Barr:1988yj,Maekawa:2003wm} offers an attractive option to break the gauge symmetry down to the SM (and further to $SU(3)_c \otimes U(1)_Q$) at the renormalizable level and by means of a quite simple Higgs sector,
namely a couple of  $SO(10)$  spinors $16_{H_{1,2}} \oplus \overline{16}_{H_{1,2}}$ and one adjoint {$45_{H}$}. 

%=====================================================================
\subsection{Hypercharge embeddings in $SO(10)\otimes U(1)_X$}
%=====================================================================

The so called flipped realization of the  $SO(10)$ gauge symmetry
requires an additional $U(1)_{X}$ gauge factor in order to provide an extra degree of freedom for the SM hypercharge identification.
For a fixed embedding of the $SU(3)_{c}\otimes SU(2)_{L}$ subgroup within $SO(10)$, the SM hypercharge can be generally spanned over the three remaining Cartans generating the abelian $U(1)^{3}$ subgroup of the $SO(10)\otimes U(1)_{X}/(SU(3)_{c}\otimes SU(2)_{L})$ coset.
There are two consistent implementations of the SM hypercharge
within the $SO(10)$ algebra (commonly denoted by standard and flipped $SU(5)$), while a third one becomes available due to the presence of $U(1)_{X}$.

In order to discuss the different embeddings
we adopt the traditional left-right (LR) basis corresponding to the $SU(3)_{c}\otimes SU(2)_{L}\otimes SU(2)_{R}\otimes U(1)_{B-L}$ subalgebra of $SO(10)$. In full generality 
one can span the SM hypercharge on the generators of $U(1)_{R}\otimes U(1)_{B-L}\otimes U(1)_{X}$:
\be
\label{YPS}
Y=\alpha T_{R}^{(3)}+\beta(B-L)+\gamma X.
\ee
The $U(1)_X$ charge is been conveniently fixed to $X_{16}=+1$ for the spinorial representation and thus $X_{10}=-2$ and also $X_{1}=+4$ for the $SO(10)$ vector and singlet, {respectively;
this is also the minimal way} to obtain an anomaly-free $U(1)_{X}$, that
allows $SO(10)\otimes U(1)_{X}$ to be naturally embedded into $E_{6}$.
It is a straightforward exercise to show that there are only three solutions
which accommodate the SM matter quantum numbers over a reducible $16 \oplus 10 \oplus 1$ representation. 
On the $U(1)_{R}\otimes U(1)_{B-L}\otimes U(1)_{X}$ bases of \eq{YPS} one obtains,
\be\label{standardabc}
\alpha=1,\ \beta=\tfrac{1}{2} ,\ \gamma=0 \, ,
\ee
which is nothing but the ``standard'' embedding of the SM matter into $SO(10)$. The second option is characterized by
\be\label{flippedSU5abc}
\alpha=-1,\ \beta=\tfrac{1}{2} ,\ \gamma=0 \, ,
\ee
which is usually denoted ``flipped $SU(5)$''~\cite{DeRujula:1980qc,Barr:1981qv}
embedding and corresponds to a sign flip of the $SU(2)_{R}$ Cartan operator $T_{R}^{(3)}$.
A third solution corresponds to
\be\label{flippedSO10abc}
\!\alpha=0 ,\ \beta=-\tfrac{1}{4} ,\ \gamma=\tfrac{1}{4} \, ,  
\ee
denoted as ``flipped $SO(10)$''~\cite{Kephart:1984xj,Rizos:1988jn,Barr:1988yj,Maekawa:2003wm} embedding of the SM hypercharge. Notice, in particular, the fundamental difference between the setting (\ref{flippedSO10abc}) with $\gamma = \tfrac{1}{4}$ and the two previous cases (\ref{standardabc}) and (\ref{flippedSU5abc}) where $U(1)_{X}$ does not play any role.

%=====================================================================
\subsection{The supersymmetric flipped $SO(10) \otimes U(1)_X$ model}
%=====================================================================

The active role of the $U(1)_{X}$ generator in the SM hypercharge
identification within the flipped $SO(10)$ scenario has relevant consequences for model building. 
In particular, the SM decomposition of the $SO(10)$ representations changes so that there are additional SM-singlets both in $16_{H}\oplus \overline{16}_{H}$ as well as in $45_{H}$.
The presence of these additional SM-singlets provides the
ground for obtaining a viable symmetry breaking with a significantly simplified renormalizable Higgs sector. 

Naively, one may guess that the pair of VEVs in $16_{H}$ (plus another conjugated pair in $\overline{16}_{H}$ to maintain the required $D$-flatness)
might be enough to break the GUT symmetry entirely, since
one component transforms as a $10$ of $SU(5)\subset SO(10)$ (cf.\! $s_H$ in Table \ref{tab:stanVSflipYukawa}), while the other one (cf.\! $n_H$ in Table \ref{tab:stanVSflipYukawa}) is identified with an $SU(5)$ singlet. 
Nevertheless, flipping is not per-se sufficient 
since the adjoint does not reduce the rank and the bi-spinor, in spite of the two qualitatively different SM-singlets involved, can lower it only by a single unit, leaving a residual $SU(5)\otimes U(1)$ symmetry.
Only when two pairs of $16_H\oplus \overline{16}_H$ (interacting via $45_H$) are introduced the two pairs of SM-singlet VEVs in the spinor multiplets may not generally be aligned and the 
little group is reduced to the SM \cite{Bertolini:2010yz}.
Given the most general renormalizable Higgs superpotential, made of the representations $45_H \oplus 16_{H_1} \oplus \overline{16}_{H_1} \oplus 16_{H_2}  \oplus \overline{16}_{H_2}$ 
\be
\label{WHFSO10}
W_H = \frac{\mu}{2} \, \Tr 45_H^2 + \rho_{ij} 16_{H_{i}} \overline{16}_{H_{j}} + \tau_{ij} 16_{H_{i}} 45_H \overline{16}_{H_{j}} \, ,
\ee
where $i,j = 1,2$, the study of the SUSY vacuum in Ref.\! \cite{Bertolini:2010yz} shows that the little group is the SM for large portions of the parameter space in which the VEVs of the 
$16_H \oplus \overline{16}_H$ pairs are not aligned. 

\renewcommand{\arraystretch}{1.3}
\begin{table}[h]
\begin{tabular}{lll}
\hline \hline
\ \
& $SO(10)$
& $SO(10)_f$
\\
\hline
$16_F$ &
$\left( D^c \oplus L \right)_{\overline{5}} \oplus \left( U^c \oplus Q \oplus E^c \right)_{10} \oplus (N^c)_1$ \quad\quad &
$\left( D^c \oplus \Lambda^c \right)_{\overline{5}} \oplus \left( \Delta^c \oplus Q \oplus S \right)_{10} \oplus (N^c)_1$
\\
\null
$10_F$ &
$\left( \Delta \oplus \Lambda^c \right)_{5} \oplus \left( \Delta^c \oplus \Lambda \right)_{\overline{5}}$ &
$\left( \Delta \oplus L \right)_{5} \oplus \left( U^c \oplus \Lambda \right)_{\overline{5}}$
\\
\null
$1_F$ &
$(S)_1$ &
$(E^c)_1$
\\
%\hline
$\vev{16_H}$ &
$\left( 0 \oplus \vev{H_d} \right)_{\overline{5}} \oplus \left( 0 \oplus 0 \oplus 0 \right)_{10} \oplus (n_H)_1$ &
$\left( 0 \oplus \vev{H_u} \right)_{\overline{5}} \oplus \left( 0 \oplus 0 \oplus s_H \right)_{10} \oplus (n_H)_1$
\\
\null
$\vev{\overline{16}_H}$ \quad\quad &
$\left( 0 \oplus \vev{H_u} \right)_{5} \oplus \left( 0 \oplus 0 \oplus 0 \right)_{\overline{10}} \oplus (n_H)_1$ &
$\left( 0 \oplus \vev{H_d} \right)_{5} \oplus \left( 0 \oplus 0 \oplus s_H \right)_{\overline{10}} \oplus (n_H)_1$
\\
\hline \hline
\end{tabular}
\mycaption{
SM decomposition of $SO(10)$ representations 
%relevant for the Yukawa sector 
in the standard (left) and flipped (right) hypercharge embedding.
A self-explanatory SM notation is used, with the outer subscripts labeling the $SU(5)$ origin.
}
\label{tab:stanVSflipYukawa}
\end{table}

Let us stress that in the flipped embedding the spinor representations include also weak doublets $H_u$ and $H_d$ that may trigger the electroweak symmetry breaking
and allow for renormalizable Yukawa interactions with the chiral matter fields distributed in the flipped embedding over a reducible $16_F \oplus 10_F \oplus 1_F$ representation. 
Notice that this matter content is needed in order to cancel the gauge anomalies of the $U(1)_X$ factor and to correctly reproduce the SM matter quantum numbers (cf.\! Table \ref{tab:stanVSflipYukawa}).

Considering for simplicity just one pair of spinor Higgs multiplets and imposing a $Z_2$ matter-parity {(negative for matter and positive for Higgs superfields) the most general Yukawa superpotential (up to $d= 5$ operators)} reads
\be
\label{YukFSO10}
W_Y =  Y_{U} 16_F 10_F 16_H 
+ \frac{1}{M_P} \left[ Y_{E} 10_F 1_F \overline{16}_H \overline{16}_H
+  Y_{D} 16_F 16_F \overline{16}_H \overline{16}_H \right] \, ,
\ee
where family indexes are understood. Notice (cf.\! Table \ref{tab:stanVSflipYukawa}) that {due to} the flipped
embedding the up-quarks receive mass at the renormalizable level,
while all the other fermion masses need Planck-suppressed effective contributions in order to achieve a realistic texture. Thus the top/bottom hierarchy is given by an $M_G / M_P \sim 10^{-2}$ factor, 
which selects naturally $\mathcal{O}(1)$ values for $\tan\beta \equiv v_u / v_d$. 
At the end, it can be shown \cite{Bertolini:2010yz} that the Yukawa superpotential in \eq{YukFSO10} can reproduce realistic textures for the SM fermions (including neutrinos), while the exotic states are automatically kept heavy by the symmetry breaking pattern.

%=====================================================================
\subsection{Minimal $E_6$ embedding}
%=====================================================================

\begin{figure}[h]
\begin{center}
\includegraphics[width=8cm]{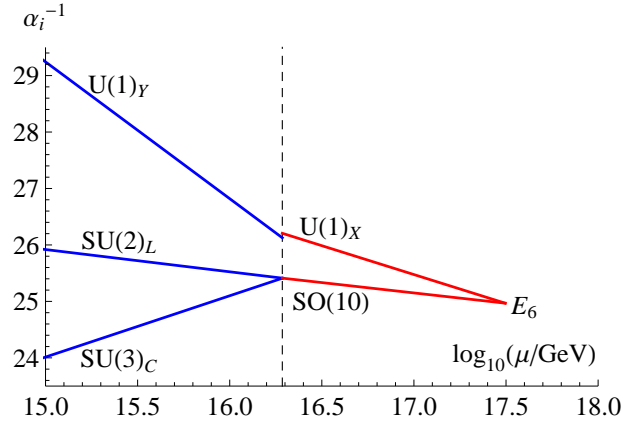}
\caption{\label{E6unification}
Sample picture of gauge unification in the {$E_6$-embedded}
$SO(10)\otimes U(1)_X$ model.
}
\end{center}
\end{figure}

The mechanism we advocate can be embedded in an underlying nonrenormalizable $E_{6}$ Higgs model featuring a pair of $27_H\oplus \overline{27}_H$ and the adjoint $78_H$.
Technical similarities apart, there is, however, a crucial difference between the $SO(10)\otimes U(1)_X$ and $E_{6}$ scenarios, that is related  to the fact that the Lie-algebra of $E_{6}$ is larger than that of $SO(10)\otimes U(1)_X$.  It has been shown long ago \cite{Buccella:1987kc} that the renormalizable SUSY $E_{6}$ Higgs model spanned over $27_H\oplus \overline{27}_H\oplus 78_H$ 
leaves an $SO(10)$ symmetry unbroken. Two pairs of $27_H \oplus \overline{27}_H$ are needed to reduce the rank by two units.
In spite of the fact that the two SM-singlet directions in the $27_H$ are exactly those of the ``flipped'' $16_H$, the little group of the $2 \times (27_H \oplus \overline{27}_H) \oplus 78_H$ 
Higgs sector remains at the renormalizable level $SU(5)$, as we explicitly show in Ref.\! \cite{Bertolini:2010yz}.
Adding nonrenormalizable adjoint interactions allows for a misalignment of the $\vev{78_H}$ from the $SU(5) \otimes U(1) \otimes U(1)$ direction, such that the little group is reduced to the SM. {Since} a one-step $E_6$ breaking with nonrenormalizable operators {is} phenomenologically problematic as mentioned earlier, we argue for a two-step breaking, via flipped $SO(10)\otimes U(1)_X$, with the $E_6$ scale near the Planck scale.
Barring detailed threshold effects,
it is interesting to see from \fig{E6unification} that 
the few percent mismatch observed within the two-loop MSSM gauge coupling evolution at the scale of the ``one-step'' grand unification is naturally accommodated in this scheme, and it is understood as an artefact of a ``delayed'' $E_{6}$ unification superseding the flipped $SO(10)\otimes U(1)_X$ partial unification.

%%%%%%%%%%%%%%%%%%%%%%%%%%%%%%%%%%%%%%%%%%%
\section{Conclusions}
%%%%%%%%%%%%%%%%%%%%%%%%%%%%%%%%%%%%%%%%%%% 
 
Longstanding results claimed that nonsupersymmetric $SO(10)$ GUTs with just the adjoint triggering the first stage of the GUT breaking cannot provide a successful gauge unification. We argued that this 
conclusion is an artefact of the tree level potential and showed that quantum corrections have a dramatic impact. In particular, a model featuring $10_H \oplus 126_H \oplus 45_H$ in the Higgs sector has all the ingredients to be a viable minimal nonsupersymmetric $SO(10)$ GUT candidate \cite{WIP}.
Analogously, supersymmetric $SO(10)$ GUTs with representations up to the adjoint do not provide a phenomenologically viable breaking to the SM. 
We pointed out that the flipped $SO(10)$ embedding offers an attractive setting for breaking the gauge symmetry directly to $SU(3)_c \otimes U(1)_Q$ at the renormalizable level, by means of a quite simple Higgs sector: $2 \times (16_H \oplus \overline{16}_H) \oplus 45_H$. The case is made for a two-step breaking of a supersymmetric $E_6$ GUT realised in the vicinity of the Planck scale via an intermediate flipped $SO(10) \otimes U(1)_X$ stage.

\end{document}